# ENHANCING DIFFUSION-WEIGHTED IMAGES (DWI) FOR DIFFUSION MRI: IS IT ENOUGH WITHOUT NON-DIFFUSION-WEIGHTED B=0 REFERENCE?

*Yinzhe Wu[1,2,3], Jiahao Huang[1,3], Fanwen Wang[1,3], Mengze Gao[2],*
*Congyu Liao[2], Guang Yang[1,3]\*, Kawin Setsompop[2]\**

[1]Department of Bioengineering and I-X, Imperial College London
[2]Department of Radiology, School of Medicine, Stanford University
[3]Cardiovascular Research Centre, Royal Brompton Hospital
\*Co-last authors

## ABSTRACT

Diffusion MRI (dMRI) is essential for studying brain microstructure, but high-resolution imaging remains challenging due to the inherent trade-offs between acquisition time and signal-to-noise ratio (SNR). Conventional methods often optimize only the diffusion-weighted images (DWIs) without considering their relationship with the non-diffusion-weighted (b=0) reference images. However, calculating diffusion metrics, such as the apparent diffusion coefficient (ADC) and diffusion tensor with its derived metrics like fractional anisotropy (FA) and mean diffusivity (MD), relies on the ratio between each DWI and the b=0 image, which is crucial for clinical observation and diagnostics. In this study, we demonstrate that solely enhancing DWIs using a conventional pixel-wise mean squared error (MSE) loss is insufficient, as the error in ratio between generated DWIs and b=0 diverges. We propose a novel ratio loss, defined as the MSE loss between the predicted and ground-truth log of DWI/b=0 ratios. Our results show that incorporating the ratio loss significantly improves the convergence of this ratio error, achieving lower ratio MSE and slightly enhancing the peak signal-to-noise ratio (PSNR) of generated DWIs. This leads to improved dMRI super-resolution and better preservation of b=0 ratio-based features for the derivation of diffusion metrics.

## 1. INTRODUCTION

Diffusion MRI (dMRI) is essential for studying brain microstructure, providing insights into tissue architecture through the measurement of water diffusion. It has become a key imaging modality in both clinical and research settings, enabling the characterization of various conditions in brain [1] and other organs [2]. Despite its utility, achieving high-resolution dMRI remains a challenge due to the inherent trade-offs between acquisition time and signal-to-noise ratio.

Several super-resolution and reconstruction methods have been proposed to improve the dMRI [3], [4], [5], [6], but they often focus on enhancing diffusion-weighted images (DWIs) individually without fully leveraging the non-diffusion-weighted (b=0) reference images, where b=0 could help enhancing the model further [7]. The b=0 image, acquired with no diffusion weighting, generally exhibits higher SNR and presents clearer structural details, because of its repeated acquisition in a dMRI session and its absence of diffusion encoding signal attenuation. The relationship between DWIs and the b=0 image is crucial for calculating diffusion metrics such as the apparent diffusion coefficient (ADC) and the diffusion tensor, which in turn are used to derive parameter maps like fractional anisotropy (FA) and mean diffusivity (MD). These parametrics are fundamental for assessing brain microstructural changes and could be sensitive to errors in the DWI-to-b=0 ratio.

Conventional deep learning approaches [8] for dMRI enhancement [9], [10] primarily utilize pixel-wise losses such as mean squared error (MSE) in DWI for model training. However, these methods fail to consider the critical ratio between DWI and b=0 images [9], [10], which is a key determinant for accurate diffusion metric calculation.

**Our Contribution: (A)** We demonstrate through a super-resolution task that, without explicitly optimizing for this ratio, the resulting enhanced DWI/b=0 ratio may exhibit significant discrepancies, which could potentially lead to inaccuracy in further derived diffusion metrics. Going further, **(B)** we proposed a novel ratio loss to relate the generated DWI with the reference b=0, maintaining the fidelity of the DWI/b=0 relationship, and thereby improving the accuracy of derived diffusion metrics. Our experiments demonstrate that incorporating the ratio loss significantly enhances the convergence behavior during training, achieving a lower ratio MSE and slightly improving the peak signal-to-noise ratio (PSNR) of the generated DWIs, providing a promising direction for advancing dMRI enhancement techniques.

## 2. METHOD

### 2.1 Problem Formulation
A typical dMRI acquisition includes repeated non-diffusion-weighted (b=0) reference image and several diffusion encoding images acquired with diffusion weightings (b-values) and several diffusion-encoding gradient directions.

For each voxel, the signal attenuation $S/S_0$ for each diffusion-weighted image (DWI) is calculated relative to the b=0 image. Here, $S_0$ is the signal intensity of the b=0 image, and $S$ is the signal intensity for each DWI image with diffusion weighting. The signal attenuation can be expressed by the Stejskal-Tanner equation as in **Eq. (1)**.

$$S/S_0 = \exp(-b \cdot \mathbf{g^T D g}), \quad (1)$$

where $b$ is the b-value, $\mathbf{g}$ is the diffusion encoding gradient direction vector, and $\mathbf{D}$ is the diffusion tensor.

Using the signal attenuation measurements $S/S_0$ from multiple DWI images with different gradient directions and b-values, we solve for the diffusion tensor in each voxel, from which further diffusion metrics (MD, FA, etc.) can be derived.

### 2.2 Motivation for $S/S_0$ Ratio-Based Validation
Enhancement in dMRI often focus on optimizing the DWI $S$ directly, without explicit consideration of the reference $S_0$. However, the ratio $S/S_0$ is the key to accurate diffusion tensor fitting and the subsequent derivation diffusion metrics. To address this, we propose incorporating the consistency of the ratio $S/S_0$ as part of the validation and evaluation objectives in dMRI enhancement tasks by its MSE distance in **Eq. (2)**.

$$d_{\text{ratio}} = \text{MSE}\left(\frac{S_{b1000}^{Result}}{S_{b0}^{GT}+\epsilon}, \frac{S_{b1000}^{GT}}{S_{b0}^{GT}+\epsilon}\right), \quad (2)$$

where $\epsilon = 1 \times 10^{-6}$ is added to prevent division by zero, and the division here refers to element-wise matrix operations.

By directly evaluating the generated DWI/b=0 ratio against the ground truth ratio, we can achieve a more realistic enhancement output that respects the fundamental diffusion relationships.

### 2.3 $S/S_0$ Ratio-log Loss for Training
Going further, we also propose the ratio-log loss for a more realistic enhancement output that respects the fundamental diffusion relationships.

By modifying the distance metric in **Eq. (2)**, we calculate the MSE loss between the logarithms of the predicted and ground truth ratios instead of the ratios themselves as in **Eq. (3)**.

$$\mathcal{L}_{\text{ratio\_log}} = \text{MSE}\left(\log\left(\frac{S_{b1000}^{Result}}{S_{b0}^{GT}+\epsilon}\right), \log\left(\frac{S_{b1000}^{GT}}{S_{b0}^{GT}+\epsilon}\right)\right) \quad (3)$$

where $\epsilon = 1 \times 10^{-6}$ is added to prevent division by zero. This approach offers several advantages:

- **Stabilized Training:** Using the logarithm helps stabilize model training, as it moderates extreme values, particularly in calculations involving divisions [11].
- **Direct Relevance to Diffusion Encoding:** In **Eq. (2)**, the original ratio-based distance metric $d_{\text{ratio}}$ measures the error in the exponential term $\exp(-b \cdot \mathbf{g^T D g})$ in **Eq. (1)**. However, by applying the logarithm in $\mathcal{L}_{\text{ratio\_log}}$, the focus shifts to the error in $\mathbf{g^T D g}$, which directly correlates with diffusion tensor fitting.

This $\mathcal{L}_{\text{ratio\_log}}$ term would be linearly added to existing loss terms with a weight of 0.01, in our case pixel-wise MSE loss $\mathcal{L}_{\text{MSE}}$ and FFT loss $\mathcal{L}_{\text{FFT}}$ [12], giving the final loss in **Eq. (4)**.

$$\mathcal{L}_{\text{total}} = 15 \times \mathcal{L}_{\text{MSE}} + 0.0025 \times \mathcal{L}_{\text{FFT}} + 0.01 \times \mathcal{L}_{\text{ratio\_log}} \quad (4)$$

## 3. EXPERIMENT SETTING

### 3.1 Dataset and Pre-processing
Pre-processed dMRI data of 13 participants from the UK Biobank [13] were used for this study. The acquisition methods and parameters detail can be found at (http://biobank.ctsu.ox.ac.uk/crystal/crystal/docs/brain_mri.pdf). The training dataset comprised their dMRI data, with a voxel size of 2 mm isotropic for DWIs (45 directions) at b=1000 s/mm$^2$ (referred as b=1000 below), complemented by 2 mm isotropic b=0 images (averaged by 5 repeats).

This study aimed to train a super-resolution model to upscale anisotropic low-resolution DWIs (voxel size: $4 \times 2 \times 2$ mm) to high-resolution (2 mm isotropic) images. To obtain the low-resolution slices downsampled volumetrically, each slice underwent anisotropic bilinear undersampling for in-plane anisotropic downsampling. The pixel values of each DWI slice with their corresponding b=0 image were normalized to be mostly within the range [0, 1] prior to model input.

In this study, we only focused on the mid and upper volumetric slices of the brain. The training set included data from 10 participants (4,800 b=1000 2D slices), while the validation set comprised 3 patients (1,440 b=1000 2D slices).

### 3.2 Model Architecture
We employed the U-Net variant in line with recent state-of-the-art generative image processing publication [14] as the backbone, where it is enhanced with multiple multi-head self-attention layers to enhance the image quality of the results produced.

### 3.3 Implementation Details
The model was trained using two NVIDIA GeForce 3090 GPUs with 24 GB memory each, employing a batch size of 50 and an initial learning rate of $1 \times 10^{-4}$, which was halved every 10 epochs. As discussed in 3.1, the input to the model were normalized 2D slices of downsampled b=1000 image, which is trained against the 2mm iso b=1000 images for output.

## 3. RESULT AND DISCUSSION

### 3.1 Does a Higher PSNR Guarantee Better Preservation of the DWI/b=0 Ratio?

Initially, we trained the U-Net model without incorporating the $\mathcal{L}_{\text{ratio\_log}}$ loss term, setting the total loss function as $\mathcal{L}_{\text{total}} = 15 \times \mathcal{L}_{\text{MSE}} + 0.0025 \times \mathcal{L}_{\text{FFT}}$. As shown in **Figure 1**, PSNR increased steadily during training; however, the $d_{\text{ratio}}$ metric initially decreased, indicating some early alignment with the DWI/b=0 relationship, but subsequently rose again as PSNR continued to improve, eventually stabilizing at a higher value.

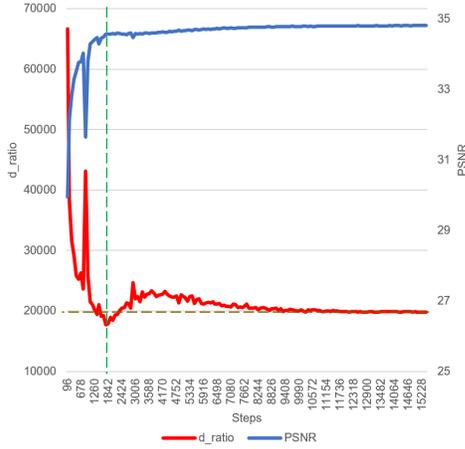

**Figure 1.** Validation curves for $d_{\text{ratio}}$ (**red**) and PSNR (**blue**) over training steps. **Red curve:** PSNR, reflecting the fidelity of individual DWI images, shows a steady monotonic increase, converging at 34.8. **Blue curve**: In contrast, $d_{\text{ratio}}$, which represents the fidelity of the DWI/b=0 ratio, initially decreases, indicating some alignment with the ground truth ratio. However, after step #1842, $d_{\text{ratio}}$ begins to rise again, eventually stabilizing at a higher value, even as PSNR continues to improve. This divergence illustrates that higher PSNR does not necessarily imply improved consistency in the DWI/b=0 ratio.

This divergence between PSNR and $d_{\text{ratio}}$ underscores a key limitation: while PSNR serves as a useful quality metric for individual DWIs, it does not adequately reflect the preservation of the critical DWI/b=0 ratio, which could affect downstream calculations such as diffusion tensor fitting.

This observation suggests that optimizing solely for $\mathcal{L}_{\text{MSE}}$ can lead to suboptimal results in diffusion metric accuracy, as $\mathcal{L}_{\text{MSE}}$ of a single DWI alone fails to capture the nuanced relationship between DWI and b=0 images essential for accurate tensor estimation. Without an explicit loss component targeting the DWI/b=0 ratio consistency, model improvements in PSNR may inadvertently degrade the integrity of this ratio, which could propagate errors in downstream diffusion analyses and compromise the reliability of derived microstructural metrics.

### 3.2 Ratio-log Loss $\mathcal{L}_{\text{ratio\_log}}$

Integrating $\mathcal{L}_{\text{ratio\_log}}$ into the loss terms (as specified in **Eq. (4)**) directs the model to consider the logarithmic DWI/b=0 ratio, effectively emphasizing $\mathbf{g}^T \mathbf{D} \mathbf{g}$, optimization as outlined in **Eq. (1)**. This approach enhances the model's ability to preserve the DWI/b=0 intensity relationship, a key factor for accurate dMRI reconstruction. As illustrated in **Figure 2**, the $d_{\text{ratio}}$ metric exhibits a consistent downward trend, despite minor fluctuations, ultimately converging at a much lower level than achieved by the baseline model without $\mathcal{L}_{\text{ratio\_log}}$ (**Table 1**). This outcome suggests that incorporating $\mathcal{L}_{\text{ratio\_log}}$ can significantly improve ratio consistency by minimizing discrepancies in the generated b=1000/b=0 relationship, which is crucial for downstream diffusion analyses.

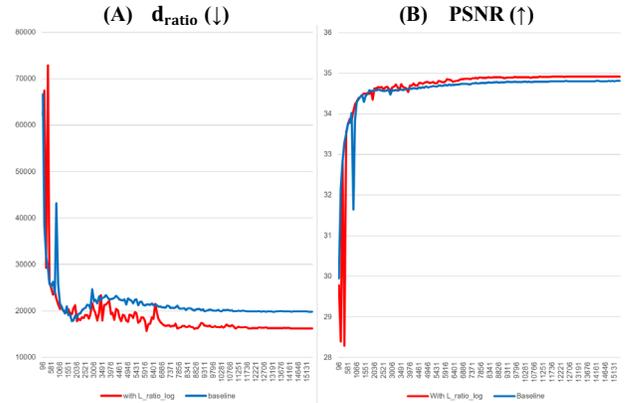

**Figure 2.** Training curves for **(A)** $d_{\text{ratio}}$ and **(B)** PSNR over epochs. In **(A)**, when $\mathcal{L}_{ratio\_log}$ is included (**red curve**) $d_{\text{ratio}}$ demonstrates a continuous decrease and converges at a lower value, highlighting improved DWI/b=0 ratio consistency. In **(B)**, PSNR shows a slight improvement with $\mathcal{L}_{ratio\_log}$ (**red curve**) compared to the baseline (**blue**), suggesting enhanced image fidelity alongside ratio preservation.

**Table 1.** Comparison of PSNR and converged $d_{\text{ratio}}$ for baseline U-Net and U-Net with $\mathcal{L}_{ratio\_log}$ added. Incorporating $\mathcal{L}_{ratio\_log}$ improves both the PSNR slightly and significantly reduces $d_{\text{ratio}}$, indicating enhanced consistency in the DWI/b=0 ratio preservation.

| U-Net | PSNR (↑) | Converged $d_{\text{ratio}}$ (↓) |
|---|---|---|
| Baseline | 34.80 | 19860 |
| With $\mathcal{L}_{ratio\_log}$ | **34.92** | **16330** |

Beyond optimizing $d_{\text{ratio}}$, the addition of $\mathcal{L}_{\text{ratio\_log}}$ also led to a slight PSNR improvement, from 34.80 to 34.92. This enhancement is particularly noteworthy, as PSNR traditionally reflects pixel-level fidelity rather than ratio consistency, where many actually studies suggested that an addition of physics informed loss could make pixel-wise metrics (e.g., PSNR) lower [15]. We hypothesize that this PSNR gain may stem from the implicit influence of the high-resolution b=0 image, which acts as a structural reference, aiding the network in aligning generated outputs with high-resolution ground truth more closely. This dual benefit suggests that $\mathcal{L}_{\text{ratio\_log}}$ not only ensures consistency in the DWI/b=0 relationship but also contributes to overall image quality, making it a valuable addition to loss function formulations in DWI super-resolution.

Even though it is expected that error in DWI/b=0 ratio would propagate to the downstream parametrics (diffusion tensor fitting, MD, FA maps etc.), this study primarily used PSNR and ratio metrics for evaluation, further validation will be done through diffusion tensor fitting and tractography to better understand the model's influence on downstream analyses.

Expanding on its utility, this loss function could be further applied to other quantitative imaging modalities that involve a reference image. By incorporating $\mathcal{L}_{\text{ratio\_log}}$ into the loss formulation, future models could jointly optimize for downstream quantitative metrics alongside image fidelity. This dual approach could enhance both the structural integrity and quantitative accuracy of reconstructed images, ensuring that crucial intensity ratios or mapping values align more precisely with ground truth references.

## 4. CONCLUSION

This study presents a novel approach to dMRI enhancement by introducing a ratio-log loss term, $\mathcal{L}_{\text{ratio\_log}}$, which explicitly optimizes the critical DWI/b=0 ratio. Our results demonstrate that incorporating $\mathcal{L}_{\text{ratio\_log}}$ not only improves the fidelity of the DWI/b=0 relationship but also contributes to a slight increase in PSNR, which is traditionally challenging to achieve with physics-informed losses. This dual benefit enhances the accuracy of diffusion metric derivations, crucial for downstream applications like diffusion tensor fitting and tractography reconstructions.

In addition, this study provides direct evidence that optimizing a single DWI without accounting b=0 may not lead to enhanced fidelity in DWI/b=0 ratio and potentially its subsequent dMRI reconstruction outcomes. Particularly, improvements in PSNR for individual DWIs do not necessarily translate to better dMRI reconstruction quality, underscoring the need for ratio-consistent loss functions relating DWI/b=0 like $\mathcal{L}_{\text{ratio\_log}}$ to ensure accurate diffusion metric derivation.

## 5. ACKNOWLEDGMENT

This research has been conducted using the UK Biobank Resource under Application Number 100203.

This work is funded in part by Imperial College London President's PhD Scholarship, in part by Imperial College London I-X Moonshot Seed Fund, and in part by UKRI Future Leaders Fellows Development Network. G. Yang's work is funded by the UKRI Future Leaders Fellowship (MR/V023799/1).

For the purpose of open access, the authors have applied a Creative Commons Attribution (CC BY) license to any Accepted Manuscript version arising.

## 7. REFERENCES

[1] D. Le Bihan, "Looking into the functional architecture of the brain with diffusion MRI," *Nat Rev Neurosci*, vol. 4, no. 6, pp. 469–480, Jun. 2003, doi: 10.1038/nrn1119.

[2] Z. Khalique, P. F. Ferreira, A. D. Scott, S. Nielles-Vallespin, D. N. Firmin, and D. J. Pennell, "Diffusion Tensor Cardiovascular Magnetic Resonance Imaging," *JACC Cardiovasc Imaging*, vol. 13, no. 5, pp. 1235–1255, May 2020, doi: 10.1016/j.jcmg.2019.07.016.

[3] J. Huang *et al.*, "Deep learning-based diffusion tensor cardiac magnetic resonance reconstruction: a comparison study," *Sci Rep*, vol. 14, no. 1, p. 5658, Mar. 2024, doi: 10.1038/s41598-024-55880-2.

[4] T. Xiang, M. Yurt, A. B. Syed, K. Setsompop, and A. Chaudhari, "DDM$^2$: Self-Supervised Diffusion MRI Denoising with Generative Diffusion Models," Feb. 2023, [Online]. Available: http://arxiv.org/abs/2302.03018

[5] D. H. J. Poot *et al.*, "Super-resolution for multislice diffusion tensor imaging," *Magn Reson Med*, vol. 69, no. 1, pp. 103–113, Jan. 2013, doi: 10.1002/mrm.24233.

[6] Q. Tian *et al.*, "SDnDTI: Self-supervised deep learning-based denoising for diffusion tensor MRI," *Neuroimage*, vol. 253, p. 119033, Jun. 2022, doi: 10.1016/j.neuroimage.2022.119033.

[7] Y. Wu *et al.*, "High-resolution reference image assisted volumetric super-resolution of cardiac diffusion weighted imaging," in *Medical Imaging 2024: Image Processing*, O. Colliot and J. Mitra, Eds., SPIE, Apr. 2024, p. 68. doi: 10.1117/12.3006008.

[8] J. Huang *et al.*, "Data- and Physics-driven Deep Learning Based Reconstruction for Fast MRI: Fundamentals and Methodologies," *IEEE Rev Biomed Eng*, pp. 1–20, 2024, doi: 10.1109/RBME.2024.3485022.

[9] S. I. Park, Y. Yim, J. Bin Lee, and H. S. Ahn, "Deep learning reconstruction of diffusion-weighted brain MRI for evaluation of patients with acute neurologic symptoms," *Sci Rep*, vol. 14, no. 1, p. 24761, Oct. 2024, doi: 10.1038/s41598-024-75011-1.

[10] E. A. Kaye *et al.*, "Accelerating Prostate Diffusion-weighted MRI Using a Guided Denoising Convolutional Neural Network: Retrospective Feasibility Study," *Radiol Artif Intell*, vol. 2, no. 5, p. e200007, Aug. 2020, doi: 10.1148/ryai.2020200007.

[11] M. Arjovsky, S. Chintala, and L. Bottou, "Wasserstein GAN," Jan. 2017.

[12] G. Yang *et al.*, "DAGAN: Deep De-Aliasing Generative Adversarial Networks for Fast Compressed Sensing MRI Reconstruction," *IEEE Trans Med Imaging*, vol. 37, no. 6, pp. 1310–1321, Jun. 2018, doi: 10.1109/TMI.2017.2785879.

[13] K. L. Miller *et al.*, "Multimodal population brain imaging in the UK Biobank prospective epidemiological study," *Nat Neurosci*, vol. 19, no. 11, pp. 1523–1536, Nov. 2016, doi: 10.1038/nn.4393.

[14] A. Nichol and P. Dhariwal, "Improved Denoising Diffusion Probabilistic Models," Feb. 2021.

[15] Z. Liu, M. Roy, D. K. Prasad, and K. Agarwal, "Physics-Guided Loss Functions Improve Deep Learning Performance in Inverse Scattering," *IEEE Trans Comput Imaging*, vol. 8, pp. 236–245, 2022, doi: 10.1109/TCI.2022.3158865.